DFT insights into MAX phase borides $Hf_2AB$ [A = S, Se, Te] in comparison with MAX phase carbides $Hf_2AC$ [A = S, Se, Te]

J. Islam[1], M. D. Islam[2], M. A. Ali[3,4,*], H. Akter[3,4], A. Hossain[2,3], M. Biswas[3,4], M. M. Hossain[3,4], M. M. Uddin[3,4], S. H. Naqib[4,5,*]

[1]*Department of Physics, Noakhali Science and Technology University, Noakhali-3814, Bangladesh*

[2]*National Institute of Textile Engineering and Research, Savar, Dhaka-1350*

[3]*Department of physics, Chittagong University of Engineering and Technology (CUET)*

*Chattogram-4349, Bangladesh*

[4]*Advanced Computational Materials Research Laboratory, Department of physics, Chittagong University of Engineering and Technology (CUET)*

*Chattogram-4349, Bangladesh*

[5]*Department of Physics, University of Rajshahi, Rajshahi 6205, Bangladesh*

## Abstract

In this work, density functional theory (DFT) based calculations were performed to compute the physical properties (structural stability, mechanical behavior, electronic, thermodynamic, and optical properties) of synthesized MAX phases $Hf_2SB$, $Hf_2SC$, $Hf_2SeB$, $Hf_2SeC$, $Hf_2TeB$, and the as-yet-undiscovered MAX carbide phase $Hf_2TeC$. Calculations of formation energy, phonon dispersion curves, and elastic constants confirmed the stability of aforementioned compounds. The obtained values of lattice parameters, elastic constants, and elastic moduli of $Hf_2SB$, $Hf_2SC$, $Hf_2SeB$, $Hf_2SeC$, and $Hf_2TeB$ showed fair agreement with earlier studies, whereas, the values of the mentioned parameters for the predicted $Hf_2TeC$ exhibit a good consequence of B replacement by C. The anisotropic mechanical properties are exhibited by the considered MAX phases. The metallic nature and its anisotropic behavior were revealed by the electronic band structure and density of states. The analysis of the thermal properties—Debye temperature, melting temperature, minimum thermal conductivity, and Grüneisen parameter—confirmed that the carbide phases were more suited than the borides phases considered herein. The MAX phase's response to incoming photons further demonstrated that they were metallic. Their suitability for use as coating materials to prevent solar heating was demonstrated by the reflectivity spectra. Additionally, this study demonstrated the impact of B replacing C in the MAX phases.

## 1. Introduction

One of the most thoroughly investigated groups of transition metal borides, carbides, or nitrides is denoted by the notation $M_{n+1}AX_n$, where M is any of the transition metal elements, A is an A-group element selected from columns 13–16 of the periodic table, and X denotes boron, carbon, or nitrogen.[1,2] The properties of metals and ceramics are combined in MAX-phase materials, which are exceptionally resistant to oxidation and corrosion, elastically rigid, have higher melting temperatures, and are lightweight and machinable. They also have excellent electrical and thermal conductivity, are resistant to thermal shock, and have plasticity at high temperatures. They can therefore be utilized as a substitute for both ceramic and metal depending on the application. MAX phase materials have recently been used for various technological and industrial purposes, including in the nuclear industry and high-temperature applications, electric circuits, sensors, heat exchangers, metal refining electrodes, and aerospace.[3–7] Researchers and engineers are encouraged to theoretically or experimentally explore the MAX phase compounds because of their appealing features. Till now, more than 150 MAX phases made up of 32 elements have been identified.[1,8]

The combination of the strong covalent M-X bond and the weak metallic M-A bond is responsible for the combined characteristics (metal and ceramic)of MAX phases. The element and composition of M, A, and X (M = Ti, V, Cr, Ta, Zr, Hf, Nb, and so on; A = Al, S, Se, Sn, As, In, Ga; and X = N, C) can readily be changed, which allows for easy modification of the MAX phase's physical and chemical characteristics. Especially, the substitution of the M and A sites are the best choice for the prediction of new MAX phases, such as (Sc, Ti, Cr, Zr, Nb, Mo, Hf, Ta)$_2$AlC,[9] (Ti, V, Nb, Zr, Hf)$_2$SnC,[10,11] (Ti, Zr, Hf)$_2$SC,[12,13] (Zr, Hf, Nb)$_2$SC,[14] $V_2$(Al, Ga)C,[15–17] $Ti_2$(Zn, Al, In, Ga)C,[18] and $Cr_2$(Al, Ge)C[19]. A mixture of A site elements also demonstrated a synergistic impact of both atoms (increased physicochemical properties) were also shown in a number of investigations. For instance, Li et al. synthesized a series of $V_2(A_xSn_{1-x})C$ phases (A = Fe, Co, Ni, Mn, or their mixture).[20] They investigated the impact of alloying A site element on the magnetic properties.[20] However, the majority of the variation in the X site was restricted to either carbon or nitrogen or a combination of the two.[14] Boron has recently been added to this list.[21] Rackl et al. have recentlysynthesized and studied the crystal structure, stability, chemical bonds, and elastic and electronic properties of prepared $Zr_2SB$ and $Hf_2$SBMAX phase borides for the first time.[21] The prepared $Hf_2$SB MAX phase exhibit metallic character and Pauli paramagnetism. Additionally, DFT calculations show that the ionic bonds are somewhat weaker than those of carbides, resulting in lower bulk moduli of this boride than for the comparablecarbide.[22] In a different work, we investigated the mechanical properties, elastic anisotropy, optical properties, dynamic stability, and thermal properties of $Hf_2$SB and $Hf_2$SC and observed a similar result for the variation of mechanical properties within these two phases.[14] Wang et al.[23] synthesized $Hf_2$SeC and compared it with $Hf_2$SC; due to a longer bonding length of Hf-Se than Hf-S, a lower bonding energy and a softer structure of $Hf_2$SeC is expected, which results in a larger thermal expansion coefficient (TEC) of $Hf_2$SeC than that of $Hf_2$SC. Similar results were also observed for $Zr_2$SeC and $Zr_2$SC.[23]Zhang et al. theorized the stability and lattice characteristics of the $Hf_2$SeB MAX phase in a different work, in which successful synthesis of this compound using the thermal explosion method in a speak plasma sintering furnace was reported.[24] Recently, $Zr_2$SeB and $Hf_2$SeB, two MAX phases that were both thermodynamically and mechanically stable, were predicted by Zhang et al.[25] They also synthesized the phases by thermal explosion technique in a spark plasma sintering furnace.

Since sulfur, selenium, and tellurium are in the same periodic table group with similar electronic structures, thus, tellurium is also expected to be a potential A-site element of the ternary carbides and/or borides; belong to the MAX phase family. Zhang et al. brought out the B-Te chemistry in light by synthesizing the first tellurium (Te) containing layered ternary compound $Hf_2$TeB using the thermal explosion method.[26] The dynamical stability and mechanical properties of $Hf_2$TeBhave also been investigated by Zhou et al.[27] Since the $Hf_2$TeB MAX phase has yet to be completely investigated, more theoretical data, such as anisotropy indices, thermo-dynamical properties, and optical properties, should be scrutinized.Motivated by the synthesis of $Hf_2$TeB and based on our previous experience of the MAX phase chemistry, we assumed that $Hf_2$TeC might be a potential member of MAX phase family. Consequently, we predict the stability of $Hf_2$TeC for the first time by calculating the formation energy, phonon dispersion curve (dynamical stability) and elastic constants (mechanical stability).

Therefore, we aimed to perform a DFT-based study for the synthesized $Hf_2$SC, $Hf_2$SB, $Hf_2$SeC, $Hf_2$SeB, and $Hf_2$TeB; and newly predicted $Hf_2$TeC. Although, the mechanical and dynamical properties of synthesized $Hf_2$TeB have been reported[27], but the mechanical anisotropy and Vickers hardness, classification of active modes in phonon dispersion, as well as thermal and optical properties are yet to be explored. Knowledge of mechanical properties is important for structural materials, but the lack of information regarding mechanical anisotropy may restrict their use, as it (anisotropy) is related to the plastic deformation, fracture propagation, and elastic instability. On the other hand, the Vickers hardness measures the ultimate strength of solids, which is pre-requisite for application is any structural components. The ultimate use of MAX phase's materials in high temperature environment (such as thermal barrier coating materials) depends on the values of parameters (Debye temperature, minimum thermal conductivity, melting temperature, Grüneisen parameter, etc.) that characterize the thermal properties of solids. Use of MAX phase materials in spaceship as coating materials to reduce solar heating depends on the value of the reflectivity. Thus, study of thermal and optical properties is scientifically important to broaden the scope of their potential uses. The aforementioned physical properties of $Hf_2TeB$, a new member of MAX phase materials, explored in this study, should also be investigated to disclose its potentials to be used in various sectors. In addition, we have also included the $Hf_2SB$,[21] $Hf_2SC$,[28] $Hf_2SeB$,[24] $Hf_2SeC$[23,28] MAX phases in this paper to bring out the Physics of replacing the A elements (and X element:B/C) on the physical properties considered herein. Thus, the structural, mechanical, electronic, thermal, optical, and mechanical properties of $Hf_2AX$ [A = S, Se, Te; X =C, B] MAX phase compounds have been thoroughly investigated in this paper.

## 2. Computational methods

The scientific community in the areas of physics and materials engineering has shown a significant deal of interest in computer-based ab-initio calculations utilizing the DFT.[29,30] The CAmbridge Serial Total Energy Package (CASTEP)[31] computer code, which is based on the DFT, was used for the crystal geometry computations and characteristics analysis in the current study of MAX phases.The generalized gradient approximation (GGA), along with Perdew−Burke−Ernzerhof (PBE),[32]was used to describethe exchange-correlation energy. The ultrasoft pseudopotential suggested by Vanderbilt[33]was chosen for calculating the electron−ion interaction. Ultrasoft pseudopotentials (USP) were introduced by Vanderbilt in 1990. The USP offers a possibility to resolve the problem of plane-wave convergence for transition metals with good accuracy. In USP, the norm-conservation constraint is relaxed and the difference in the charge densities calculated from the all-electron and pseudo-orbitals is described in terms of a small number of localized augmentations functions. This allows calculations to be performed with the lowest possible cutoff energy for the plane-wave basis set.The Broyden−Fletcher−Goldfarb−Shanno (BFGS) scheme[34] was used to ensure geometry optimization.The pseudo-atomic calculations were performed, taking only valence electrons to reduce core-electron effects.The crystal geometry was optimized and the physical properties were calculated using a k-point of 10×10×3 (Monkhorst−Pack schemes),[35] and a cutoff energy of 500 eV. The convergence criteria were selected as follows: total energy: $5 \times 10^{-6}$ eV/atom; maximum force: 0.01 eV/Å; maximum ionic displacement: $5 \times 10^{-4}$ Å; maximum stress: 0.02 GPa.

## 3. Results and discussion

### 3.1. Structural Properties and Stability

Figure 1 depicts the crystal structure of ternary $Hf_2SB$ MAX Phases which crystallize to a hexagonal structure [space groupP63/mmc] like the other C or N-containing MAX compounds.[14,21–23,25,27,28,36] The investigated MAX phase compounds have two formula units, each of which has four atoms.The atomic positions of these MAX phase compounds are as follows: Hf atoms at 4*f* (0.3333, 0.6667, $Z_M$), A atoms at 2*d*(0.3333, 0.6667, 0.75), and X atoms at 2*a* (0, 0, 0) sites. The other crystallographic data of $Hf_2SB$ (as a representative) are presented in supplementary.

**Table 1.** Calculated lattice parameters (*a* and *c*), internal parameter ($z_M$), *c*/*a* ratio, and formation energy of $Hf_2AX$ [A = S, Se, Te; X =C, B] MAX phase compounds.

| Phase | *a* (Å) | % of deviation | *c* (Å) | % of deviation | $z_M$ | *c/a* | $E_F$ (eV/atom) | References |
|---|---|---|---|---|---|---|---|---|
| $Hf_2SB$ | 3.506 | 1.13 | 12.229 | 1.03 | 0.603 | 3.487 | -1.26 | This study |
| | 3.467 | | 12.104 | | 0.604 | 3.491 | | Expt.[21] |
| | 3.482 | 0.43 | 12.137 | 0.27 | 0.603 | 3.485 | | Theo.[21] |
| | 3.506 | 1.13 | 12.159 | 0.45 | 0.603 | 3.468 | | Theo.[14] |
| | 3.508 | 1.20 | 12.222 | 0.97 | 0.603 | 3.483 | | Theo.[27] |
| $Hf_2SC$ | 3.300 | 2.0 | 11.757 | 2.1 | 0.099 | 3.562 | -1.13 | This study |
| | 3.369 | | 12.017 | | 0.600 | 3.566 | | Expt.[21] |
| | 3.360 | | 11.997 | | | 3.570 | | Expt.[22] |
| | 3.424 | 1.63 | 12.184 | 1.39 | 0.600 | 3.558 | | Theo.[14] |
| | 3.369 | 0.01 | 11.993 | 0.19 | 0.101 | 3.560 | | Theo.[23] |
| $Hf_2SeB$ | 3.552 | 0.85 | 12.559 | | 0.099 | 3.535 | -1.16 | This study |
| | 3.522 | | 12.478 | | 0.598 | 3.542 | | Expt.[27] |
| | 3.550 | 0.79 | 12.573 | | 0.599 | 3.541 | | Theo.[27] |
| $Hf_2SeC$ | 3.470 | 1.42 | 12.514 | 0.99 | 0.095 | 3.605 | -1.03 | This study |
| | 3.422 | | 12.391 | | | 3.621 | | Expt.[25] |
| | 3.436 | 0.41 | 12.452 | 0.49 | | 3.624 | | Theo.[25] |
| | 3.438 | 0.47 | 12.501 | 0.89 | | 3.636 | | Theo.[28] |
| $Hf_2TeB$ | 3.629 | 0.68 | 13.355 | 1.74 | 0.591 | 3.679 | -0.86 | This study |
| | 3.604 | | 13.126 | | | 3.641 | | Expt.[36] |
| $Hf_2TeC$ | 3.559 | | 13.236 | | 0.088 | 3.718 | -0.69 | This study |

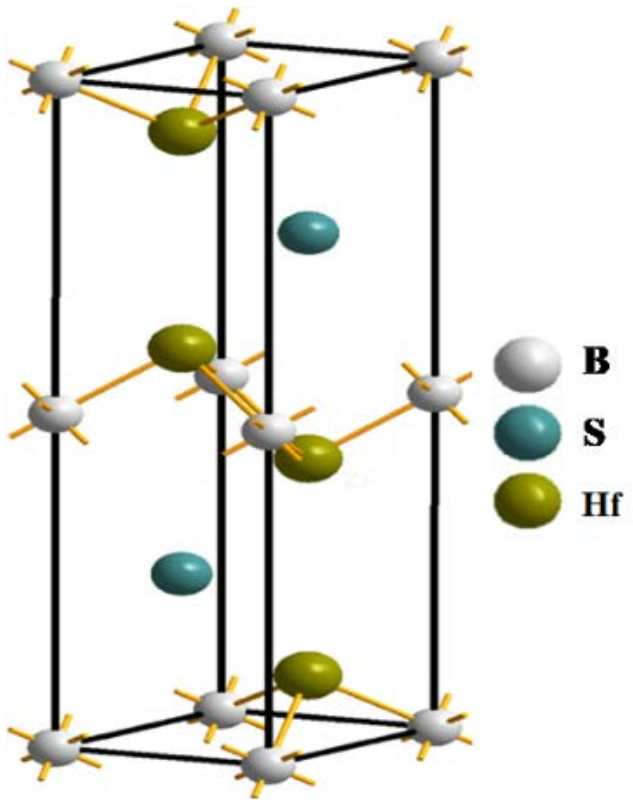

**Figure 1.** Crystal structure of the $Hf_2SB$ compound.

The DFT-optimized lattice constants (*a*, *c*), *c/a* ratios, and internal parameters ($z_M$) of the studied MAX boride and carbide phases are listed in Table 1, together with other theoretical and available experimental results. These DFT-optimized lattice constants (*a* and *c*) show good consistency with earlier experimental as well as theoretical studies, as can be shown in Table 1. Lattice constant *c* of the $Hf_2TeB$ phase is determined to have a maximum divergence from the experimental value of 1.74%. These agreements suggest that the current investigation is highly reliable. The $Hf_2TeC$ MAX phase is predicted for the first time; as a result, it was not possible to make a comparison with other previous reports for this particular carbide. This DFT-predicted result for $Hf_2TeC$ can be a useful reference work for designing future experimental as well as theoretical research works.

The phonon dispersion curve (PDC) can be used to forecast a compound's structural stability, thermodynamic stability, and vibrational contribution.[37] As most of the MAX phases are synthesized at a higher thermal treatment (at a higher temperature), those materials may not be stable under all conditions. In order to check the dynamic stability of $Hf_2TeX$ (X = B,C), phonon dispersion curves (PDCs) have been calculatedusing the Density Functional Perturbation Theory (DFPT) with linear-response method.[14,38] The calculated phonon dispersion curves (PDCs) and total phonon density of states (PHDOS) of individual MAX phase compounds along the high symmetry paths of the Brillouin zone (BZ) are displayed in Figure 2. The presence of only positive frequencies indicates the dynamical stability of the compound. As there is no negative frequencies are detected in the PDCs of $Hf_2TeC$, as displayed in Figure 2 (e), this MAX phase compound is predicted to be dynamically stable like its boride counterpart $Hf_2TeB$ (presented in Figure 2 (f)). Dynamical stability is also observed for the other MAX phases and presented in Figure 2 (a-d).

To further aid in comprehending the bands, the PHDOS of $Hf_2TeB$ and $Hf_2TeC$ are shown beside the PDCs. Figure 2 shows that all of the MAX phase carbides have a distinct gap between their transverse optical (TO) and longitudinal optical (LO) frequencies, but no gaps are evident for the MAX phase borides. The separation between LO (top) and TO (bottom) modes at the central G point is 2.2, 1.6, and 1.6 THz for $Hf_2SC$, $Hf_2SeC$, and $Hf_2TeC$, respectively. Furthermore, for all the MAX phases including carbides and borides, there is no phononic bandgap between the acoustic and low-energy optical phonon modesbut there is a pronounced phononic bandgap between high-energy and low-energy optical modes. Most importantly, at the Brillouin zone center point (G) of the PDCs, the frequency of acoustic modes is zero in all the MAX phases under study, which is another indication of the structural stability of the MAX phases under consideration.

The lattice dynamics of crystal is especially fascinating in terms of zone-center phonon modes. The 8 atoms that make up the 211 MAX phases give them a total of 24 phonon branches or vibration modes. There are 21 optical modes, and the remaining three are zero-frequency acoustic modes at the Γ -point. Six of these 21 optical modes are IR-active, seven are Raman-active, and the other eight are silent modes. The optical phonon modes with a Brillouin zone-center can be expressed by followings groups:

$$\Gamma_{opt.} = 2A_{2u} + 4E_{1u} + 2E_{1g} + 4E_{2g} + A_{1g} + 2B_{1u} + 2B_{2g} + 4E_{2u}$$

where IR active modes are $A_{2u}$ and $E_{1u}$, Raman active modes are $A_{1g}$, $E_{1g}$, and $E_{2g}$ and silent modes are $B_{1u}$, $B_{2g}$ and $E_{2u}$.[39,40]

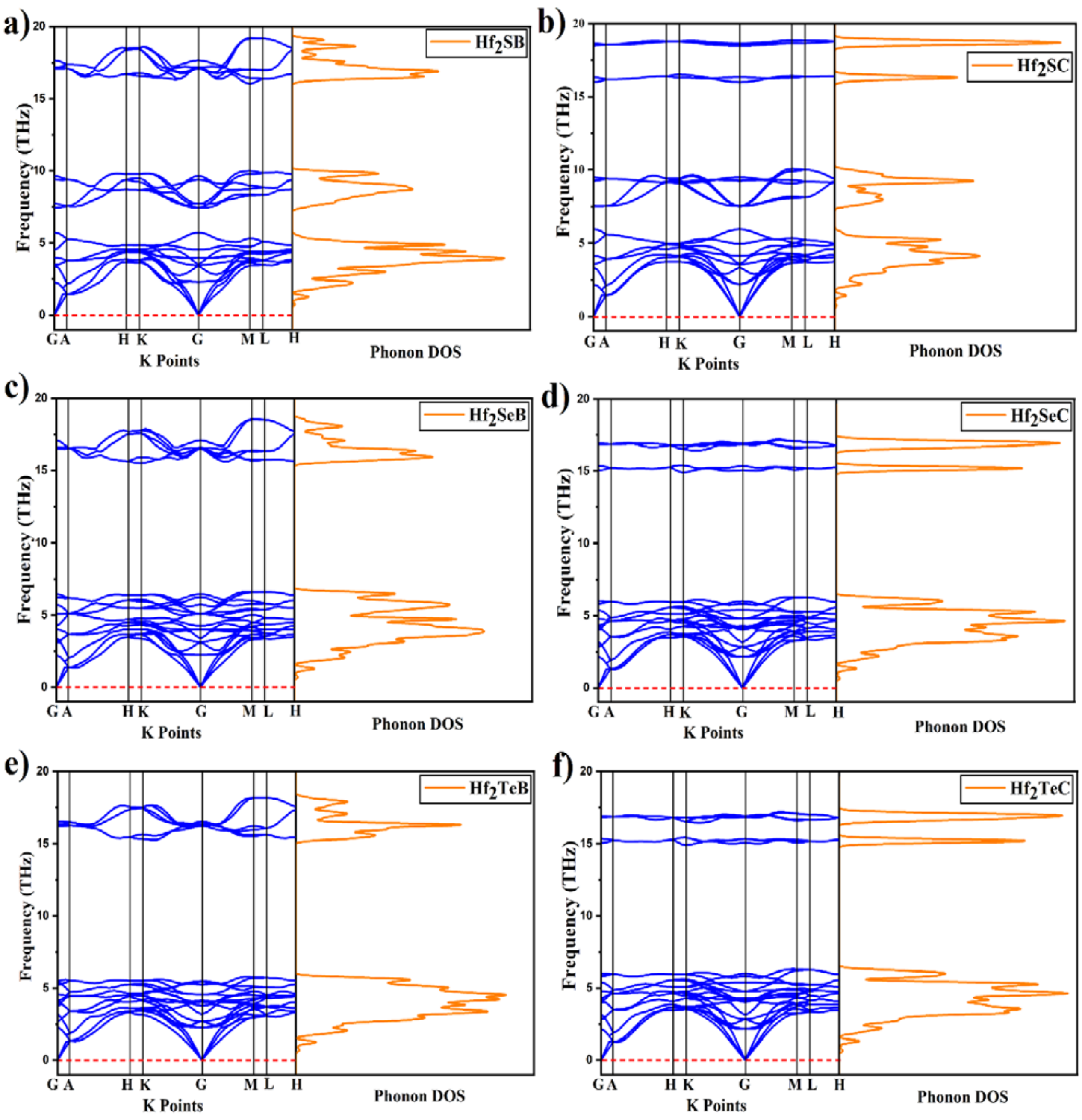

**Figure 2.** The PDCsandPHDOSof a) $Hf_2SB$, b) $Hf_2SC$, b) $Hf_2SeB$, d) $Hf_2SeC$, e) $Hf_2TeB$, and f) $Hf_2TeC$ compounds. The dashed line (red) is at zero phonon frequency.

The obtained modes for the $Hf_2AX$ [A = S, Se, Te; X =C, B] phases in this study are in agreement with earlier theoretical investigations of the different 211 MAX phases. A specific vibrational frequency corresponds to each mode. Degenerate modes are those that occasionally have two or more modes with the same frequency but cannot be distinguished from one another.

For this reason, Table S4 contains six IR active modes and seven Raman active modes for each phase.

### 3.1.1. Mechanical properties

**Table 2.** Calculated stiffness constants ($C_{ij}$), bulk modulus ($B$), machinability index ($B/C_{44}$), Cauchy pressure (CP), shear modulus ($G$), Young's modulus ($Y$), Poisson's ratio (ν) and Pugh ratio (G/B) of $Hf_2AX$ [A = S, Se, Te; X =B, C] MAX phase compounds.

| **Parameters** | **$Hf_2SB$** | **$Hf_2SC$** | **$Hf_2SeB$** | **$Hf_2SeC$** | **$Hf_2TeB$** | **$Hf_2TeC$** |
|---|---|---|---|---|---|---|
| $C_{11}$(GPa) | 285 | 350 | 232 | 287 | 182 | 257 |
| | 286[a] | 344[d] | 243[f] | 308[e] | 206[c] | |
| | 285[b] | 311[b] | 244[c] | 287[g] | | |
| | 265[c] | 330[e] | | | | |
| $C_{33}$ (GPa) | 298 | 371 | 274 | 305 | 243 | 277 |
| | 296[a] | 369[d] | 281[f] | 314[e] | 232[c] | |
| | 298[b] | 327[b] | 281[c] | 306[g] | | |
| | 305[c] | 345[e] | | | | |
| $C_{44}$(GPa) | 130 | 176 | 120 | 130 | 107 | 112 |
| | 122[a] | 175[d] | 119[f] | 135[e] | 107[c] | |
| | 130[b] | 149[b] | 119[c] | 130[g] | | |
| | 125[c] | 150[e] | | | | |
| $C_{12}$(GPa) | 81 | 98 | 98 | 77 | 77 | 81 |
| | 79[a] | 116[d] | 90[f] | 93[e] | 71[c] | |
| | 81[b] | 97[b] | 90[c] | 77[g] | | |
| | 100[c] | 100[e] | | | | |
| $C_{13}$(GPa) | 92 | 118 | 88 | 97 | 83 | 92 |
| | 84a | 138[d] | 97[f] | 105[e] | 85[c] | |
| | 92[b] | 121[b] | 87[c] | 98[g] | | |
| | 90[c] | 118[e] | | | | |
| $C_{55}$(GPa) | 130 | 176 | 120 | 130 | 107 | 112 |
| $C_{66}$(GPa) | 101 | 126 | 67[c] | 104 | 52[c] | 87 |
| | | | 79[f] | | | |
| $B$ (GPa) | 155 | 193 | 143 | 157 | 120 | 146 |
| | 151[a] | 204[d] | 144[f] | 158[g] | 125[c] | |
| | 156[b] | 181[b] | 144[c] | | | |
| $B/C_{44}$ | 1.19 | 1.09 | 1.19 | 1.21 | 1.12 | 1.30 |
| | 1.23[a] | 1.16[d] | 1.21[f] | 1.22[g] | 1.17[c] | |
| | 1.20[b] | 1.21[b] | 1.21[c] | | | |
| *CP* (GPa) | -49 | -78 | -22 | -53 | -30 | -31 |
| | -49[b] | -52[b] | -29 | -53[g] | -36 | |
| $G$ (GPa) | 111 | 142 | 89 | 112 | 73 | 97 |
| | 111[a] | 134[d] | 95[f] | 113[g] | 81[c] | |
| | 112[b] | 120[b] | 95[c] | | | |
| $Y$ (GPa) | 268 | 344 | 222 | 273 | 183 | 238 |
| | 267[a] | 330[d] | 233[f] | 273[g] | 199[c] | |
| | 270[b] | 295[b] | 233[c] | | | |
| ν | 0.21 | 0.20 | 0.24 | 0.21 | 0.24 | 0.22 |
| | 0.20[a] | 0.23[d] | 0.23[f] | 0.21[g] | | |
| | 0.21[b] | 0.23[b] | | | | |
| *G/B* | 0.71 | 0.73 | 0.62 | 0.71 | 0.61 | 0.66 |
| | 0.73[a] | 0.65[d] | 0.66[f] | 0.72[g] | 0.65[c] | |
| | 0.71[b] | 0.66[b] | 0.66[c] | | | |
| [a]Ref-[21], [b]Ref-[14], [c]Ref-[27], [d]Ref-[22], [e]Ref-[23], [f]Ref-[25], [g]Ref-[28] | | | | | | |

It is essential to study the elastic properties of a material to understand its mechanical behaviors, such as mechanical stability, ductility/brittleness, hardness, and machinability. The elastic constants ($C_{ij}$) are key parameters in figuring out the mechanical properties of solids. Because they are connected to various solid-state phenomena, including inter-atomic bonding and phonon spectra, elastic constants of solids are very important. Elastic constants are also relatedto the Debye temperature, thermal expansion, and specific heat from a thermodynamic perspective. Therefore, before moving on to a material's applications, it is highly enlightening to look into its elastic properties. In this study, we have explored the detailed elastic properties of $Hf_2SB$, $Hf_2SC$, $Hf_2SeB$, $Hf_2SeC$, $Hf_2TeB$, and $Hf_2TeC$. The elastic properties of $Hf_2SB$, $Hf_2SC$, $Hf_2SeB$, $Hf_2TeB$, and $Hf_2SeC$ have been investigated earlier[14,21–23,25,27,28] and here we have rechecked for further validationof our calculations. To the best of our knowledge, the elastic properties of $Hf_2TeC$ have not yet been documented. Moreover, the layered ternary carbide phase $Hf_2TeC$ is not yet synthesized. This study might be useful for designing and carrying out experimental and theoretical research on $Hf_2TeC$ as well as other unexplored MAX phase compounds. The calculated elastic properties with previous reports have been listed in Table 2. The hexagonal system possesses five independent elastic constants; $C_{11}$, $C_{33}$, $C_{44}$, $C_{12}$, and $C_{13}$. For a hexagonal system, the compound is said to be mechanically stable if it satisfies the stability conditionsas presentedbythe equation S2.

From Table 2, it can be observed that the studied MAX phase compounds are mechanically stable. This study provides the first demonstration of the mechanical stability of the $Hf_2TeB$ and $Hf_2TeC$ MAX phases using elastic constants. This prediction can be helpful for the experimental synthesis of unexplored ternary carbide MAX phase $Hf_2TeC$. Moreover, these elastic constants offer important details regardingstiffness in different directions. The elastic constants $C_{11}$ and $C_{33}$ depict the stiffness against applied stress along [100] and [001] directions, respectively. From Table 2, it is revealed that the value of $C_{33} > C_{11}$ for all the studied MAX phases, which indicates that the $c$-axis has stronger deformation resistance than the $a$-axis. The fact that $C_{44}$ is less than $C_{33}$ and $C_{11}$ suggests that shear deformation is less difficult than axial deformation. The unequal values of $C_{11}$, $C_{33}$, and $C_{44}$ ($C_{11} \neq C_{33} \neq C_{44}$) imply different atomic arrangements and hence different bonding strengths along $a$ and $c$-axis, and shear planes. $C_{12}$ and $C_{13}$ have lesser values than the other three elastic constants. Additionally, $C_{12}$ and $C_{13}$ have smaller values compared to those of the other three elastic constants. The stiffness constants $C_{12}$ and $C_{13}$ indicate applied stress towards the $a$-axis along with uniaxial strain along the $b$-, and $c$-axis, respectively. With the exception of the $C_{12}$ value for $Hf_2SeB$ and $Hf_2SeC$, it is interesting to notice that the stiffness constant values of the carbide MAX phases are bigger than the corresponding boride MAX phases, indicating that the carbide phase is stiffer than its boride phase. Table 2 further shows that the stiffness constants in $Hf_2SC$ are larger than those in other MAX phases; while those in $Hf_2TeB$ are lower. This result suggests that among the studied compounds, $Hf_2SC$ is the hardest MAX phase and $Hf_2TeB$ is the softest MAX phase. The $C_{ij}$ of the predicted phases, $Hf_2TeC$, should be compared with those of other Hf-based 211 carbides/borides. The $C_{11}\{C_{33}[C_{44}]\}$is found for $Hf_2TeC$ is 257{277[112]} GPa, which is higher than those of $Hf_2SeB$ [Table 2], $Hf_2TeB$ [Table 2], $Hf_2AlB$ [199 GPa],[9] $Hf_2SiB$ [223 GPa],[41] $Hf_2PB$ [254 GPa][41,42]. It is required to study the brittleness/ductility of material fora particular device application. The $C_{ij}$ values help to predict the ductile/brittle behavior of a material. The Cauchy pressure ($CP$), which is defined as the difference between the elastic constant values $C_{12}$ and $C_{14}$, is positive or negative depending on whether it represents the nature of ductility or brittleness.[14,43] The negative value of $CP$ indicates thebrittle behavior of thetitled compounds. This result has been checked further by calculating the Pugh ratio and Poisson's ratio values using the computed elastic moduli. The Voigt-Reuss-Hill (VRH) approximation schemes[44,45] are widely used to calculate the three different elastic moduli; bulk modulus ($B$), shear modulus ($G$), and Young's modulus ($Y$). The equations needed to calculate these elastic moduli are given in the supplementary file (equations S3).[1,46] The higher (lower) value of $B$ and other elastic moduli implies thehard (soft) nature of a material.[25,47] From the tabulated values of $B$, $G$, and $Y$, it can be noticed that $Hf_2TeB$ is softer and $Hf_2SC$ is harder compared to other MAX phase compounds. This result strongly supports the stiffness constants calculations. The machinability index ($\mu_{\mathrm{M}}$),[48] is a useful parameter to predict the mechanical engineering performance of a material which can be obtained by using the value of $B$ and $C_{44}$ with the formula, $\mu_{\mathrm{M}} = \frac{B}{C_{44}}$. The higher (lower) value of $\mu_{\mathrm{M}}$ signifies greater ease (difficulty) to make a desired shape of a solid. The highest value of $\mu_M$ is observed for $Hf_2TeC$ and the lowest value for $Hf_2SC$. The ductile/brittle behavior of the studied MAX phases has been investigated further using the Pugh ratio ($G/B$) and Poisson's ratio. The required equation to calculate the Poisson's ratio is expressed by the equation S4. The critical value of the brittle-ductile nature for thePugh ratio ($G/B$) is 0.571,[49] and it is 0.26[50] for the Poisson's ratio ($v$). The compound shows ductile (brittle) behavior, which possesses Pugh and Poisson's ratio values above (below) this critical line. The studied MAX phases are all brittle in nature, which has been also revealed from the *CP* calculation.

### 3.1.2. Mulliken Population Study

The atomic population analysis can be used to determine the charge transfer mechanism in a compound. As seen in Table S1, the negative charge of the atom indicates the electron receiver and the atom's positive charge indicates theelectron donor during compound formation. Moreover, charge is transferred fromHf to S and B/C for $Hf_2SB$ and $Hf_2SC$ MAX phases. Furthermore, in the case of other MAX phases, both M (Hf) and A (Se/Te) exhibited positive charge indicating the electron donor nature of M (Hf) and A (Se/Te) atoms. For those MAX phases, the electron is transferred from M (Hf) and A (Se/Te) to X (B/C) atomic species. It can be confirmed from the periodic nature of the electronic configurations that the metallic nature is increased from top to bottom in the same series of periodic table atoms. The charge transfer mechanism also confirms the existence of ionic behavior of the MAX phases.

The Mulliken bond overlap population (BOP) indicates the bonding and anti-bonding nature of the individual bond in the compound by the positive and negative values, respectively. The BOP value of zero indicates no significant interaction between the atoms involved in the bonding, while the higher value of BOP represents a higher degree of covalency of the bond.Table S2 shows a strong covalent bond between Hf-X (B, C) for all the MAX phase compounds. Here it should be noted that the degree of covalency of the Hf-B bond is higher than that of the Hf-C bond, resulting in the higher bonding strength of the Hf-B bond. In conclusion, it can be predicted from the Mulliken population analysis that all MAX phases exhibited both ionic and covalent bonding nature.

### 3.1.3. Vickers Hardness

In section 3.1.1, we have discussed several mechanical parameters including stiffness constants ($C_{ij}$), bulk modulus ($B$), shear modulus ($G$), and Young's modulus ($Y$) of individual MAX phase compounds. But the intrinsic hardness of a material is completely different from those selected mechanical parameters though there may be a correlation among them.[51] For this reason, in this study, we calculate the hardness (Vickers hardness) of all $Hf_2AX$ [A = S, Se, Te; X = C, B] compounds using the formula given in equation S7. The obtained results from the calculations are summarized in Table 3.

From Table S2, the Mulliken bond overlap population confirmed that the Hf-X (X = B/C) covalent bonds are stronger than Hf-A (A = S/Se) in MAX phase compounds.[14] Moreover, it has been found from Table S2 that, Hf-B bonds are stronger than Hf-C for all MAX phases which is an indication of enhanced mechanical properties of MAX phase borides compared to the carbides.[37,51] The Vickers hardness (Table 3) of $Hf_2AX$ (A = Se and Te) borides are higher than $Hf_2AX$ (A = Se and Te) carbides as predicted. The Vickers hardness of $Hf_2SC$, on the other hand, is higher than that of $Hf_2SB$. In this study, the calculated Vickers hardness values can be ranked as follows: $Hf_2SC$ (10.30 GPa) > $Hf_2SB$ (7.75 GPa) > $Hf_2SeB$ (3.56 GPa) > $Hf_2SeC$ (3.27 GPa) > $Hf_2TeB$ (3.05 GPa) > $Hf_2TeC$ (2.79 GPa).

**Table 3.** Calculated Mulliken bond number $n^{\mu}$, bond length $d_{\mu}$, bond overlap population $P^{\mu}$, metallic population $P^{\mu'}$, bond volume $V_b^{\mu}$, bond hardness $H_v^{\mu}$ of μ-type bond and Vickers hardness $H_v$ of $Hf_2AX$ [A = S, Se, Te; X =B,C] MAX phase compounds.

| **Compounds** | **Bond** | $n^{\mu}$ | **dμ(Å)** | $P^{\mu}$ | $P^{\mu'}$ | $V_b^{\mu}$ **(Å$^3$)** | $H_v^{\mu}$ **(Gpa)** | $H_v$ **(Gpa)** |
|---|---|---|---|---|---|---|---|---|
| $Hf_2SB$ | B-Hf | 4 | 2.390 | 1.44 | 0.01497 | 13.331 | 14.07 | 7.746 |
| | S-Hf | 4 | 2.700 | 0.81 | | 19.221 | 4.265 | |
| $Hf_2SC$ | C-Hf | 4 | 2.236 | 1.26 | 0.00929 | 10.781 | 17.59 | 10.295 |
| | S-Hf | 4 | 2.600 | 0.92 | | 16.949 | 6.025 | |
| $Hf_2SeB$ | B-Hf | 4 | 2.403 | 1.76 | 0.014 | 34.323 | 3.563 | 3.563 |
| $Hf_2SeC$ | C-Hf | 4 | 2.335 | 1.48 | 0.007 | 32.632 | 3.270 | 3.270 |
| $Hf_2TeB$ | B-Hf | 4 | 2.4277 | 1.79 | 0.01502 | 38.0875 | 3.046 | 3.046 |
| $Hf_2TeC$ | C-Hf | 4 | 2.363 | 1.51 | 0.00757 | 36.306 | 2.792 | 2.792 |

### 3.1.4. Elastic Anisotropy

Elastic anisotropy of a solid is important to study to explore the direction dependent bonding natures along different crystallographic directions. The knowledge of anisotropy indices provides significant information regarding possible microcracks under stress and unusual phonon modes. Therefore, a detailed exploration of anisotropy indices is required for practical applications of a material. The anisotropy factors of $Hf_2SB$ and $Hf_2SC$ have been investigated earlier.[14] The detailed anisotropy indices of $Hf_2SeB$, $Hf_2SeC$, $Hf_2TeB$, $Hf_2TeC$ are yet to be investigated. Considering this, we have calculated the various anisotropy factors of these MAX phases along different crystallographic directions and tabulated those in Table S3.

The three different shear anisotropy indices ($A_1$, $A_2$, and $A_3$) have been obtained using the well-known relations as presented in the supplementary file (equations S11). A value of $A_i$ (i = 1, 2, 3) equal to 1 indicates isotropic nature, otherwise anisotropic. The non unit (1) values of $A_i$ indicate the anisotropic behavior of the titled compounds. The bulk modulus anisotropy factors along the *a*- and *c*-axes, indicated as $B_a$ and $B_c$,are found out using the relations as presented in supplementary file (equations S12).

The unequal values of $B_a$ and $B_c$ ($B_a \neq B_c$) indicate anisotropic nature. The observed values of $B_a$ and $B_c$ for $Hf_2SB$ and $Hf_2SC$ show notable deviation with previous study, though other anisotropy factors show good agreement.[14]The ratio of linear compressibility coefficients ($k_c/k_a$) is calculated using the relation, given by equation S13.

The calculated values of $k_c/k_a$ ratio listed in Table S3 deviates from unity, further ensuring anisotropic nature of the studied MAX phases. The bulk modulus anisotropy factor ($A_B$) and the shear modulus factor anisotropy ($A_G$) are also estimated using the equations S14.

The universal anisotropic index ($A^U$) is another indicator that is taken into account while analyzing anisotropy. The equation S15 is used to get the $A^U$ from the Voigt and Reuss approximated *B* and *G*.

The estimated values of $A_B$, $A_G$, and $A^U$are listed in Table S3. The non-zero values of $A_B$, $A_G$, and $A^U$ notify the anisotropic nature of the studied MAX phases. From this study of different anisotropy factors, it can be concluded that all the investigated MAX phase compounds are anisotropic in nature, and consequently their mechanical/elastic properties arealso direction dependent.

### 3.2. Electronic Band Structure (EBS) and Density of States (DOS)

The knowledge of the electronic band structure (EBS) of a compound is required to understand its optical and electronic transport properties. The energy band structures of the MAX phases are calculated along the high-symmetry points in the k-space directions such as G–A–H–K–G–M–L–H within the BZ and are depicted in Figure S1. The horizontal red dashed line at zero of the energy scale indicates the Fermi level ($E_F$). Figure S1 shows that the conduction and valance bands overlap at the $E_F$ for the studied phases, indicating their metallic nature. The electronic band structure reveals anisotropic conductivity since the dispersion curves in the basal plane and *c*-direction differ significantly from one another. Similar anisotropic conductivity for other MAX phases has been also reported in previous studies.[22,28,52]

The total and partial density of states (DOS) is investigated to get a clear perception of the electronic properties of the MAX phases under investigation. From Figure 3, it is obvious that the total DOS at the $E_F$ is finite, indicatingthe metallic feature of all the MAX phases. Moreover, relative heights or the existence of a pseudo gap in the TDOS at $E_F$ indicate the presence of a stable covalent bond (stability of MAX phases) within the crystal structure.[13] The obtained TDOS at $E_F$ for $Hf_2SB$, $Hf_2SC$, $Hf_2SeB$, $Hf_2SeC$, $Hf_2TeB$, and $Hf_2TeC$ are 3.5, 1.33,[28] 3.5, 1.8, 3.7, and 1.7 states per eV, respectively. Thus, the change in the A (A = S, Se, or Te) site element does not significantly affect the DOS value at the $E_F$, however the TDOS for boride-based MAX phases is significantly higher than for carbide-based MAX phases, indicating that $Hf_2AB$s are more conductive than the $Hf_2AC$s compounds.

For more theoretical understanding, the DOS of individual atoms Hf, A (S, Se, Te), and X (B, C) in the MAX phase structures are calculated and presented in Figure 3. The DOS curves have been shown from -10 eV to 5 eV in order to better comprehend the electronic states of each atom at various energy levels.The TDOS peaks at lower energy range from -10 eV to -2 eV are mainly contributed by Hf-*p*, Hf-*d*, S-*p*, and C-*p* states. The Hf-*p*, Hf-*d*, and B-*p* states are largely responsible for the emergence of DOS in the valance band near $E_F$(~ 0 eV).

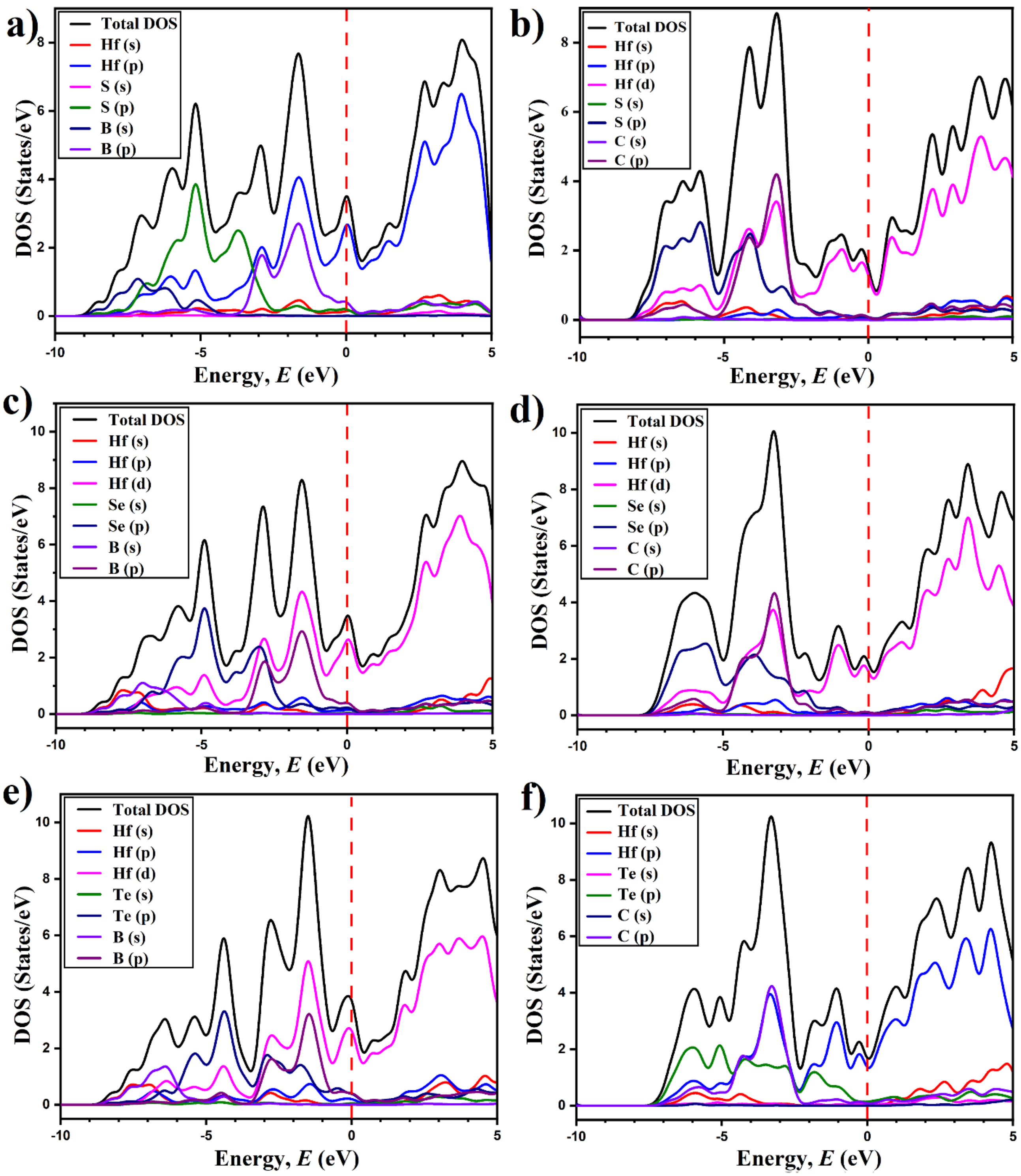

**Figure 3.** TDOS and partial density of states (PDOS) of a) $Hf_2SB$, b) $Hf_2SC$, b) $Hf_2SeB$, d) $Hf_2SeC$, e) $Hf_2TeB$, and f) $Hf_2TeC$ compounds.

### 3.3. Thermal properties

The MAX phase materials have the potential to be used in a variety of high-temperature technologies, including effectivethermal barrier coatings, heating elements, and electrical connections.[8] Therefore, in order to reveal their uses in various high-temperature technologies, a thorough investigation of the thermodynamic characteristics of MAX phase materials is required. Consequently, fundamental thermodynamic parameters such as Debye temperature ($\Theta_D$), minimum thermal conductivity ($K_{min}$), melting temperature ($T_m$) and Grüneisen parameter (γ) have been explored in this study for the $Hf_2SB$, $Hf_2SC$, $Hf_2SeB$, $Hf_2SeC$, $Hf_2TeB$, and $Hf_2TeC$ MAX phases. The required equations for calculating $\Theta_D$, $T_m$, $K_{min}$, and γ are embedded in the supplemental section (equations S16-S19). Previously, the thermodynamic properties of $Hf_2SB$,[14,21] $Hf_2SC$,[14,22] and $Hf_2SeC$[28] have been reported. As far as we are aware, there is still no information available about the thermodynamic properties of the $Hf_2SeB$, $Hf_2TeB$, and $Hf_2TeC$ MAX phases.

The Debye temperature is connected to the lattice vibration, thermal expansion coefficient, specific heat, and melting point of thesolid. It is termed as the maximum frequency mode of vibration in a solid. The calculation of $\Theta_D$ using elastic moduli is specified as one of the standard approaches.[53,54] The values of transverse sound velocity, $v_t$, longitudinal sound velocity, $v_l$, and averaged sound velocity, $v_m$ are used to estimate $\Theta_D$ using.

The calculated values of$\rho$, $v_t$, $v_l$, $v_m$, $\Theta_D$, $K_{min}$, γ, and $T_m$ are listed in Table 4. The carbide phases have higher values of $\Theta_D$ compared to their boride counterparts as shown in Table 4. The calculated values of $\Theta_D$ for $Hf_2SB$ and $Hf_2SC$ show good agreement with earlier calculations. In the case of $Hf_2SeC$, the present estimation of the $\Theta_D$ value exhibits deviation to a notable extent from the previously reported one, as displayed in Table 4.

**Table 4.** Calculated crystal density, longitudinal, transverse and average sound velocities ($v_l$, $v_t$, $v_m$), Debye temperature $\Theta_D$, minimum thermal conductivity $K_{min}$, Grüneisen parameter γ and melting temperature $T_m$ of $Hf_2AX$ [A = S, Se, Te; X =B, C] MAX phase compounds.

| **Phases** | **$\rho$ (g/cm$^3$)** | **$v_l$(m/s)** | **$v_t$(m/s)** | **$v_m$(m/s)** | **$\Theta_D$(K)** | **$K_{min}$ (W/mK)** | **γ** | **$T_m$(K)** |
|---|---|---|---|---|---|---|---|---|
| $Hf_2SB$ | 10.196 | 5451 | 3299 | 3647 | 428 | 0.78 | 1.32 | 1656 |
| Ref-[21] | | | | | 426 | | | |
| Ref-[14] | 10.198 | 5465 | 3309 | 3657 | 430 | 0.79 | 1.32 | 1656 |
| $Hf_2SC$ | 12.005 | 5643 | 3439 | 3799 | 470 | 0.90 | 1.28 | 1960 |
| Ref-[14] | 10.192 | 5784 | 3431 | 3800 | 454 | 0.85 | 1.60 | 1778 |
| Ref-[22] | 11.290 | 5820 | 3442 | 3813 | 463 | | | |
| $Hf_2SeB$ | 10.805 | 4921 | 2870 | 3198 | 367 | 0.66 | 1.45 | 1461 |
| $Hf_2SeC$ | 11.395 | 5185 | 3135 | 3465 | 406 | 0.74 | 1.32 | 1672 |
| Ref-[28] | | | | | 490 | 0.86 | | 1674 |
| $Hf_2TeB$ | 10.797 | 4486 | 2600 | 2886 | 321 | 0.55 | 1.45 | 1264 |
| $Hf_2TeC$ | 11.354 | 49247 | 2923 | 3237 | 366 | 0.64 | 1.36 | 1540 |

The temperature at which a solid substance starts to melt is defined as melting temperature. Knowing a material's melting temperature is necessary for high-temperature applications. The melting temperature is estimated using elastic constants (elastic constants $C_{11}$ and $C_{33}$ are connected to uniaxial stress) with the help of well-known expression as presented in the supplementary file (equation S17). Interesting to note from Table 4 that the carbide phase shows a higher melting temperature compared with its boride counterpart. This trend is also revealed in the Debye temperature calculation. The possible reason behind this behavior is that the carbide phases exhibit a higher value of elastic constants than their boride counterparts. Among the investigated MAX phases, $Hf_2SC$ has the highest melting temperature while $Hf_2TeB$ has the lowest.

It is essential to disclose the ability of a material to conduct heat energy for thermal device applications. The knowledge of minimum thermal conductivity, $K_{min}$ is crucial for the high-temperature application of a material. The equation S18 was used to compute $K_{min}$. The maximum value of $K_{min}$ is observed for $Hf_2SC$ and the minimum value for $Hf_2TeB$. Similar to $\Theta_D$ and $T_m$ calculations, carbide phases reveal a higher value of $K_{min}$compared with their boride counterparts. These results suggest that carbide MAX phases can be more suitable over boride counterparts for high-temperature technologies.

The Grüneisen parameter (γ) is a significant indicator for analyzing anharmonic effects within the crystal. It was calculated using the Poisson’s ratio by using equation S19. According to Table 4, the estimated values of γ are seen to remain within the range of 0.85–3.53, as is typical for polycrystalline materials within the range of 0.05–0.46.53.[55] The present studied MAX phases exhibit low anharmonic effects as the values of γ close to the lower limit. It is also interesting to note that compared to their carbide counterparts, the boride phases exhibit a relatively higher anharmonic effect.

**3.4. Optical properties**

The material response to incident photon energy and, consequently, electronic properties can be well understood by exploring the optical functions. The optical functions of $Hf_2SB$ and $Hf_2SC$ MAX phases have been studied earlier.[14] To the best of our survey, the optical functions of $Hf_2SeB$, $Hf_2SeC$, $Hf_2TeB$, and $Hf_2TeC$ MAX phases still need to be explored. To ascertain whether these various carbide and boride MAX phases are appropriate for particular application, this study thoroughly explores them.In this section, important optical functions such as the optical absorption ($\alpha$), photoconductivity ($\sigma$), reflectivity ($R$), and real ($\varepsilon_1$) and imaginary ($\varepsilon_2$) parts of dielectric functions have been depicted in Figure 4as a function of photon energy.

Figure 4 (a) shows the optical absorption profile of the studied MAX phases. The absorption coefficient ($\alpha$) evaluates a material's incident energy-absorbing ability. A higher value of the absorption coefficient indicates a higher absorbing capability of the material and vice-versa. From Figure 4 (a), we can notice that the optical absorption initiates at zero photon energy for all the MAX compounds, which increases with increasing photon energy. The maximum absorption peaks for these MAX phases are noticed within 7.0 to 15 eV of photon energies, which indicates that these compounds are good candidates for effective absorbing materials within this energy range (ultraviolent region). The $Hf_2SC$ shows significantly higher optical absorption than its boride counterpart, $Hf_2SB$. The absorption nature is approximately similar for other carbides and borides counterparts except for a little difference in the spectra positions and heights of the peaks.

The optical conductivity reflects the results of optical absorption analysis. Figure 4 (b) illustrates the optical conductivity profile of the studied MAX phases. Additionally, for all MAX phases, optical conductivity begins at zero photon energy, supporting the metallic character of all MAX compounds. This result strongly supports the optical absorption and electronic band structure calculations of the studied MAX phases. A notable optical conductivity peak is noticed in these MAX compounds' infrared to near-visible regions. Some other sharp conductivity peaks are also observed in the 5 to 10 eV of photon energies for all the phases. The $Hf_2SC$ exhibits larger conductivity spectra compared with its Boride counterpart, $Hf_2SB$. All other borides and carbides show approximately similar behavior with their corresponding counterpart. This result also justifies the analysis of the optical absorption profile.

Figure 4 (c) and 4 (d) display the real ($\varepsilon_1$) and imaginary ($\varepsilon_2$) parts of dielectric functions, respectively. Figure4 (c) exhibits the negative value of the real part of dielectric functions for all the MAX phases at zero and in the low energy region, which is an expected behavior for a metallic system. The imaginary part of the dielectric function is directly associated with the optical absorption profile, and the imaginary part tends to have approximately zero value in the very high energy region. The $Hf_2SB$ shows the negligible value of the imaginary part above 12 eV, and all other MAX phases show above 20 eV, suggesting these MAX phases can be used as transparent materials in the high energy region. The reflectivity profile provides useful information about the suitability of a material as a reflector in a practical device. From Figure 4 (e), it can be observed that the reflectivity starts at zero photon energy with a value of 0.79, 0.71, 0.64, 0.62, 0.58, and 0.56 for $Hf_2SC$, $Hf_2SB$, $Hf_2TeB$, $Hf_2SeB$, $Hf_2SeC$, and $Hf_2TeC$ MAX phases, respectively. The maximum reflectivity peak is observed at about 12 eV of photon energy for $Hf_2SB$ as this boride phase shows lower absorption spectra near about this energy than other MAX phases. Very high reflectivity (above 90%) of $Hf_2SB$ in the energy range from 11 eV to 14 eV in the ultraviolet (UV) region suggests that this particular MAX phase boride can be used selectively as an excellent reflector of the UV radiation in this range.

The refractive index, $n(\omega)$, and extinction coefficients, $k(\omega)$ of the studied MAX phases have been displayed in Figure 4 (f) and (g), respectively. The $n(\omega)$ regulates the propagation velocity of electromagnetic radiation through a material, whereas $k(\omega)$ determines the amount of radiation attenuation while traversing the material. The static values of $n(0)$ are 17.2, 11.7, 8.9, 8.4, 7.4, and 7.0 for $Hf_2SC$, $Hf_2SB$, $Hf_2TeB$, $Hf_2SeB$, $Hf_2SeC$, and $Hf_2TeC$. The static values of $n(0)$ for the studied MAX phases follow the same order with the static value of reflectivity at zero photon energy, i.e., $Hf_2SC > Hf_2SB > Hf_2TeB > Hf_2SeB > Hf_2SeC > Hf_2TeC$. The $k\,(\omega)$ curves are observed to vary approximately similarly to the $\varepsilon_2(\omega)$ as both these parameters are related to the optical absorption.

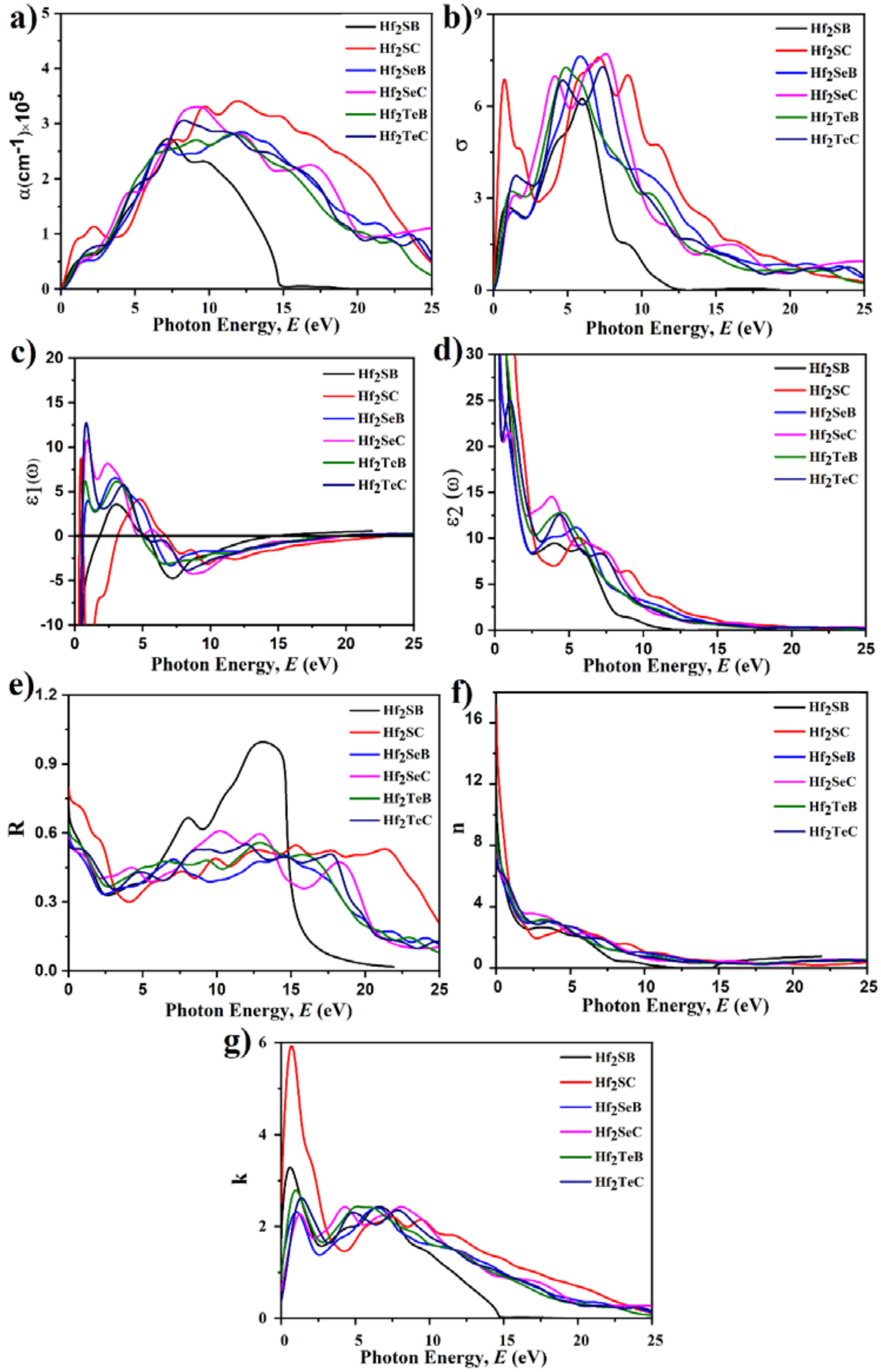

**Figure 4.**Calculated (a) absorption coefficient, $\alpha$, (b) optical conductivity, $\sigma$, (c)real part, $\varepsilon_1$ of dielectric function, (d) imaginary part, $\varepsilon_2$ of dielectric function, (e) reflectivity, $R$, (f) refractive index, $n$, and (g) extinction coefficient, $k$ of $Hf_2AX$ (A = S, Se and Te; X = B and C) MAX phases as a function of photon energy.

## 4. Conclusions

Using the DFT-based CASTEP code, the physical characteristics of the MAX phases $Hf_2TeB$ and $Hf_2TeC$ have been calculated and compared to those of $Hf_2SB$, $Hf_2SC$, $Hf_2SeB$, and $Hf_2SeC$. The Te-containing carbide MAX phase $Hf_2TeC$ has been predicted for the first time in this study. According to the various stability requirements, the predicted MAX carbide phase ($Hf_2TeC$) is elastically and lattice dynamically stable as well as possesses negative energy of formation. Therefore, $Hf_2TeC$ might be synthesized like the other stable and synthesized MAX phases. The investigation of lattice parameters, elastic constants, and elastic moduli on $Hf_2SB$, $Hf_2SC$, $Hf_2SeB$, $Hf_2SeC$, and $Hf_2TeB$ reveals good consistency with previous reports. The analysis of elastic constants reveals that the carbide phase $Hf_2SC$ is the stiffest, and the boride phase $Hf_2TeB$ is the softest MAX phase among the investigated materials. Among the examined MAX phases, $Hf_2TeC$ has the highest machinability index and $Hf_2SC$ has the lowest value, indicating that $Hf_2TeC$ is the easiest phase to shape-machine. Since all of the MAX phases under investigation in this study exhibit anisotropy as determined by various anisotropy factors, they all display direction-dependent elastic and mechanical characteristics.When compared to their boride counterparts, all of the carbide phases exhibit greater values for melting temperature, minimum thermal conductivity, and Debye temperature. The larger value of the elastic constants in the carbide phases compared to their boride counterparts may be the cause of this trend. The maximum value of melting temperature is found for $Hf_2SC$ (1960 K), and the lowest value for $Hf_2TeB$ (1264 K) among the studied MAX phases. All of the MAX phases have low anharmonic characters, whereas the borides exhibit higher anharmonic effect compared to their carbide counterpart. The study of optical functions, electronic band structure, and density of states ensures the metallic nature of all the MAX phases under consideration. The reflectivity curve starts with 0.79, 0.71, 0.64, 0.62, 0.58, and 0.56 for $Hf_2SC$, $Hf_2SB$, $Hf_2TeB$, $Hf_2SeB$, $Hf_2SeC$, and $Hf_2TeC$ MAX phases, respectively at zero photon energy. The accuracy of the current DFT-based comprehensive research is strongly supported by the obtained values of the various physical parameters, which show good agreement with earlier theoretical and experimental results where available. The authors of this publication have high hopes that this research will aid in the designing of new MAX phase materials and exploring their intriguing features for applications in practical devices from different perspectives.

**Declaration of interests**

The authors declare that they have no known competing financial interests or personal relationships that could have appeared to influence the work reported in this paper.

**Acknowledgments**

This work was carried out with the aid of a grant (grant number: 21-378 RG/PHYS/AS_G - FR3240319526) from UNESCO-TWAS and the Swedish International Development Cooperation Agency (SIDA). The views expressed herein do not necessarily represent those of UNESCO-TWAS, SIDA or its Board of Governors.

**CRediT Author contributions**

**J.Islam:** Writing – original draft**. M.D.Islam:** Writing – original draft**. M.A.Ali:** Conceptualization, Methodology, Formal analysis, Validation, Project administration, Writing – original draft, Supervision. **H. Akter**: Plotting figures, Data calculations. **A. Hossain**: Plotting figures, Data calculations **M. Biswas**: Plotting figures, Data calculations. **M.M. Hossain:** Writing – review & editing, Validation. **M.M. Uddin:** Writing – review & editing, Validation. **S.H. Naqib:**Formal analysis, Writing – review & editing, Validation.

# Electronic Supplementary Information

## DFT insights into the MAX phase borides $Hf_2AB$ [A = S, Se, Te] in comparison with MAX phase carbides $Hf_2AC$ [A = S, Se, Te]

J. Islam[1], M. D. Islam[1], M. A. Ali[2,3,*], H. Akter[2,3], A. Hossain[2,3], M. Biswas[2,3], M. M. Hossain[2,3], M. M. Uddin[2,3], S. H. Naqib[3,4,*]

[1]*National Institute of Textile Engineering and Research, Savar, Dhaka-1350*

[2]*Department of physics, Chittagong University of Engineering and Technology (CUET) Chattogram-4349, Bangladesh*

[3]*Advanced Computational Materials Research Laboratory, Department of physics, Chittagong University of Engineering and Technology (CUET) Chattogram-4349, Bangladesh*

[4]*Department of Physics, University of Rajshahi, Rajshahi 6205, Bangladesh*

**Corresponding Authors:** ashrafphy31@cuet.ac.bd, salehnaqib@yahoo.com

**Equations used to investigate various properties of $Hf_2AX$ [A = S, Se, Te; X = C, B]**

The information about the synthesizability of $Hf_2AX$ MAX phases, can be known calculating the formation energy using the equation as presented below:[1]

$$E_{for}^{Hf_2AX} = \frac{E_{total}^{Hf_2AX} - (xE_{solid}^{Hf} + yE_{solid}^{A} + zE_{solid}^{X})}{x+y+z}\ ;\ \mathrm{A = S, Se, Te;\ X = C, B} \qquad \text{(S1)}$$

Here $E_{total}^{Hf_2AX}$ indicates the total energy of the compound after performing optimization of the unit cell. $E_{solid}^{Hf}$ , $E_{solid}^{A}$ , $E_{solid}^{X}$ indicate energy of the single-phase Hf, A, X respectively. The parameters *x*, *y*, *z* represents the number of atoms in the unit cell for Hf, A, and X respectively. In this case, $x = 4$, y = 2, and z = 2.

For hexagonal system, the compound is said to be mechanically stable if it satisfies the following stability conditions: [2]

$C_{11} > 0$, $C_{33} > 0$, $C_{11}$-$C_{12} > 0$, $C_{44} > 0$, $(C_{11} + C_{12})C_{33} - 2(C_{13})^2 > 0$ (S2)

The three different elastic moduli Young's modulus (*Y*), bulk modulus (*B*), and shear modulus (*G*), are estimated using the values of the elastic constants with the help of the formulae given below:[3–5]

$$B = (B_V + B_R)/2 \quad \text{and} \quad G = (G_V + G_R)/2 \qquad \text{(S3)}$$

$$\text{Where, } B_V = [2(C_{11} + C_{12}) + C_{33} + 4C_{13}]/9;$$

$$B_R = C^2/M; C^2 = (C_{11} + C_{12})C_{33} - 2C_{13}^2; M = C_{11} + C_{12} + 2C_{33} - 4C_{13};$$

$$G_V = (M + 12C_{44} + 12C_{66})/30; G_R = \left(\frac{5}{2}\right)[C^2C_{44}C_{66}]/[3B_VC_{44}C_{66} + C^2(C_{44} + C_{66})]$$

$$\text{and } Y = 9BG/(3B + G)$$

Where, $B$ and $G$ indicate bulk modulus and shear modulus, respectively.

The Poisson's ratio ($v$) is calculated using the formula:

$$v = \frac{3B - 2G}{6B + 2G} \tag{S4}$$

The Mulliken charge of an atom ($\alpha$) is found out using the formula as presented below:[6]

$$Q(\alpha) = \sum_k w_k \sum_{\mu}^{on\ \alpha} \sum_{v} P_{\mu v}(k)\, S_{\mu v}(k) \tag{S5}$$

and the bond population associated withtwo atoms α and β is calculated by the equation:[6]

$$P(\alpha\beta) = \sum_k w_k \sum_{\mu}^{on\ \alpha} \sum_{v}^{on\ \beta} 2P_{\mu v}(k)\, S_{\mu v}(k) \tag{S6}$$

where $P_{\mu v}$, and $S_{\mu v}$ stand for the density matrix elements, and the overlap matrix.

The bond harness (Vickers hardness) of the MAXHf$_2$AX [A = S, Se, Te; X = C, B] phases is found out using the equation:[7]

$$H_v^{\mu} = 740(P^{\mu} - P^{\mu'})(v_b^{\mu})^{(-5/3)} \tag{S7}$$

Where, the symbols $P^{\mu}$, and$P^{\mu'}$ indicate Mulliken overlap population, and metallic population, respectively. The metallic population of a unit cell of volume $V$ can be estimated using the total number of free electrons through the following equation:

$$n_{free} = \int_{E_P}^{E_F} N(E)dE \text{as} P^{\mu'} = n_{free}/V \tag{S8}$$

$E_P$ indicates pseudo-gap energy, $v_b^{\mu}$ stands for volume of the $\mu$-type bond, which is calculated using the bond length $d^{\mu}$ of the $\mu$-type bond, and the bond number $N_b^{v}$ of ν-type per unit volume through the equation:

$$v_b^{\mu} = (d^{\mu})^3 / \sum_{v} [(d^{\mu})^3 N_b^{v}] \tag{S9}$$

The hardness of a complex multiband compound can be estimated by taking into account the geometric average of all bond harnesses:[8,9]

$$H_V = \left[\Pi^{\mu}\left(H_v^{\mu}\right)^{n^{\mu}}\right]^{1/\Sigma n^{\mu}} \tag{S10}$$

where, $n^{\mu}$ implies the bonds number of $\mu$-type bond in the multiband compound.

The shear anisotropy factors for different planes (100), (010), and (001) are evaluated using the following expression:[10]

$$A_1 = \frac{\frac{1}{6}(C_{11} + C_{12} + 2C_{33} - 4C_{13})}{C_{44}}$$

$$A_2 = \frac{2C_{44}}{C_{11} - C_{12}} \tag{S11}$$

$$A_3 = A_1.A_2 = \frac{\frac{1}{3}(C_{11} + C_{12} + 2C_{33} - 4C_{13})}{C_{11} - C_{12}}$$

The bulk modulus anisotropy factor along the *a*-and *c*-directions can be determined via following equations:[11]

$$B_a = a\frac{dP}{da} = \frac{\Lambda}{2+\alpha},$$

$$B_c = c\frac{dP}{dc} = \frac{B_a}{\alpha}, \tag{S12}$$

where $\alpha = \frac{(C_{11}+C_{12})-2C_{13}}{C_{33}+C_{13}}$and $\Lambda = 2(C_{11} + C_{12}) + 4C_{13}\alpha + C_{33}\alpha^2$

The linear compressibility ratio ($k_c/k_a$), is calculatedvia the following equation:[11]

$$\frac{k_c}{k_a} = \frac{C_{11} + C_{12} - 2C_{13}}{C_{33} - C_{13}} \tag{S13}$$

The bulk modulus and shear modulus anisotropy parameters $A_B$ and $A_G$ are calculated using the following equations:[11]

$$A_B = \frac{B_V - B_R}{B_V + B_R} \times 100\%$$

$$A_G = \frac{G_V - G_R}{G_V + G_R} \times 100\% \tag{S14}$$

The universal anisotropic factor ($A^U$) is analyzed using the Voigt and Reuss approximated *B* and *G* through the following equation:[12]

$$A^U = 5\frac{G_V}{G_R} + \frac{B_V}{B_R} - 6 \geq 0 \tag{S15}$$

The Debye temperature ($\Theta_D$) is estimated with the help of well-known Anderson method,[13] which uses the value of average sound velocity, $v_m$:

$$\Theta_D = \frac{h}{K_B}\left[\left(\frac{3n}{4\pi}\right)\frac{N_A\rho}{M}\right]^{\frac{1}{3}} v_m$$

$$v_m = \left[\frac{1}{3}\left(\frac{1}{{v_l}^3} + \frac{2}{{v_t}^3}\right)\right]^{-\frac{1}{3}} \qquad \text{(S16)}$$

$$v_l = \left(\frac{3B + 4G}{3\rho}\right)^{\frac{1}{2}}$$

$$v_t = \left(\frac{G}{\rho}\right)^{\frac{1}{2}}$$

where, $v_l$, $v_t$, $\rho$, $N_A$, $M$, n, $K_B$ stand for longitudinal sound velocity, transversesound velocity, density, Avogadro's number, molecular mass, number of atoms in the unit cell, Boltzmann's constant, and Planck's constant.

The melting temperature is found out via the following expression:[14]

$$T_m = 354 + 4.5\ \frac{2C_{11} + C_{33}}{3} \qquad \text{(S17)}$$

Theminimum thermal conductivity, $K_{min}$, is determined via the following equation:[11]

$$K_{min} = K_B v_m \left(\frac{M}{n\rho N_A}\right)^{-\frac{2}{3}} \qquad \text{(S18)}$$

The Grüneisen parameter (γ) is a crucial parameter associated with anharmonic effects existing within the crystal, which isevaluated by the use of the Poisson's ratio via following expression:

$$\gamma = \frac{3(1 + \upsilon)}{2(2 - 3\upsilon)} \qquad \text{(S19)}$$

**Table S1.**The crystallographic data of $Hf_2SB$

| Lattice constants (Å) | | Element | Atoms | Fractional Coordinates | | |
|---|---|---|---|---|---|---|
| | | | | *x* | *y* | *z* |
| *a* | 3.4671 | | | | | |
| *b* | 3.4671 | B | 1 | 0.0 | 0.0 | 0.0 |
| *c* | 12.1046 | B | 2 | 0.0 | 0.0 | 0.5 |
| Cell Angles (Degrees) | | S | 1 | 0.3333 | 0.6667 | 0.25 |
| α | 90 | S | 2 | -0.3333 | -0.6667 | 0.75 |
| β | 90 | Hf | 1 | 0.3333 | 0.6667 | 0.6047 |
| γ | 120 | Hf | 2 | -0.3333 | -0.6667 | 1.1047 |
| Cell Volume (Å$^3$) | | Hf | 3 | 0.6667 | 0.3333 | -0.6047 |
| *V* | 126.012553 | Hf | 4 | -0.6667 | -0.3333 | -1.1047 |

**Table S2.** Mulliken atomic and bond overlap populations (BOP) of $Hf_2AX$ [A = S, Se, Te; X = C, B] MAX phase compounds.

| Phases | Atomic population | | | | | | Bond overlap population | | |
|---|---|---|---|---|---|---|---|---|---|
| | **Atoms** | **s** | **p** | **d** | **Total** | **Charge (e)** | **Bond** | **Bond number $n^{\mu}$** | **Bond overlap population $P^{\mu}$** |
| $Hf_2SB$ | B | 1.24 | 2.58 | 0.00 | 3.82 | -0.82 | B-Hf | 4 | 1.44 |
| | S | 1.82 | 4.60 | 0.00 | 6.42 | -0.42 | S-Hf | 4 | 0.81 |
| | Hf | 0.42 | 0.14 | 2.83 | 3.39 | 0.62 | | | |
| $Hf_2SC$ | C | 1.54 | 3.39 | 0.00 | 4.93 | -0.92 | C-Hf | 4 | 1.26 |
| | S | 1.80 | 4.63 | 0.00 | 6.43 | -0.43 | S-Hf | 4 | 0.92 |
| | Hf | 0.67 | -0.34 | 2.99 | 4.00 | 0.68 | | | |
| $Hf_2SeB$ | B | 1.22 | 2.59 | 0.00 | 3.82 | -0.82 | B-Hf | 4 | 1.76 |
| | Se | 0.28 | 4.52 | 0.00 | 5.34 | 0.66 | Se-Hf | 4 | -0.32 |
| | Hf | 0.59 | 0.50 | 2.83 | 3.92 | 0.08 | | | |
| $Hf_2SeC$ | C | 1.54 | 3.33 | 0.00 | 4.87 | -0.87 | C-Hf | 4 | 1.48 |
| | Se | 0.99 | 4.52 | 0.00 | 5.51 | 0.49 | Se-Hf | 4 | -0.05 |
| | Hf | 0.52 | 0.46 | 0. 84 | 3.81 | 0.19 | | | |
| $Hf_2TeB$ | B | 1.22 | 2.58 | 0.00 | 3.80 | -0.80 | B-Hf | 4 | 1.79 |
| | Te | 0.71 | 4.27 | 0.00 | 4.97 | 1.03 | Te-Hf | 4 | -0.40 |

| | | | | | | | | |
|---|---|---|---|---|---|---|---|---|
| | Hf | 0.60 | 0.65 | 2.87 | 4.11 | 0.11 | | | |
| $Hf_2TeC$ | C | 1.54 | 3.32 | 0.00 | 4.86 | -0.86 | C-Hf | 4 | 1.51 |
| | Te | 0.89 | 4.28 | 0.00 | 5.16 | 0.84 | Te-Hf | 4 | -0.04 |
| | Hf | 0.51 | 0.59 | 2.88 | 3.99 | 0.01 | | | |

**Table S3.** Calculated values of anisotropy indices ($A_1$, $A_2$, $A_3$, $B_a$, $B_c$, $k_c/k_a$, $A_G$, $A_B$, and $A^U$) of $Hf_2AX$ [A = S, Se, Te; X = C, B] MAX phases compounds.

| **Phases** | $\boldsymbol{A_1}$ | $\boldsymbol{A_2}$ | $\boldsymbol{A_3}$ | $\boldsymbol{k_c/k_a}$ | $\boldsymbol{B_a}$ | $\boldsymbol{B_c}$ | $\boldsymbol{A_G}$ | $\boldsymbol{A_B}$ | $\boldsymbol{A^U}$ |
|---|---|---|---|---|---|---|---|---|---|
| $Hf_2SB$ | 0.76 | 1.27 | 0.97 | 0.88 | 392.6 | 841.4 | 0.78 | 0.06 | 0.08 |
| Ref-[15] | | | | 0.93 | | | | | |
| Ref-[16] | 0.76 | 1.27 | 0.97 | 0.88 | 447.2 | 506.2 | 0.78 | 0.06 | 0.08 |
| $Hf_2SC$ | 0.67 | 1.39 | 0.94 | 0.83 | 480.9 | 1109.3 | 1.55 | 0.13 | 0.16 |
| Ref-[16] | 0.65 | 1.39 | 0.90 | 0.81 | 505.5 | 627.3 | 1.71 | 0.16 | 0.17 |
| Ref-[17] | | | | 0.80 | | | | | |
| $Hf_2SeB$ | 0.73 | 1.79 | 1.30 | 0.82 | 354.3 | 832.8 | 3.47 | 0.15 | 0.36 |
| $Hf_2SeC$ | 0.75 | 1.23 | 0.93 | 0.81 | 354.0 | 837.2 | 0.76 | 0.18 | 0.08 |
| Ref-[18] | 0.75 | 1.24 | 0.93 | 0.81 | 443.0 | 548.0 | | | 0.08 |
| $Hf_2TeB$ | 0.64 | 2.0380 | 1.3111 | 0.58 | 276.7 | 970.1 | 5.41 | 0.98 | 0.59 |
| $Hf_2TeC$ | 0.77 | 1.2727 | 0.9924 | 0.83 | 363.1 | 870.1 | 0.78 | 0.12 | 0.08 |

**Table S4.**Theoreticalwave numbers ω and symmetry assignment of the IR-active and Raman-active modes of $Hf_2AX$ [A = S, Se, Te; X = C, B] MAX phases compounds.

| Theoretical Mode symmetry | Activity ($cm^{-1}$) | | | | | | | | | | | | | | | | |
|---|---|---|---|---|---|---|---|---|---|---|---|---|---|---|---|---|---|
| | IR $Hf_2SB$ | Raman $Hf_2SB$ | Theoretical Mode symmetry | IR $Hf_2SC$ | Raman $Hf_2SC$ | Theoretical Mode symmetry | IR $Hf_2SeB$ | Raman $Hf_2SeB$ | Theoretical Mode symmetry | $Hf_2SeC$ | $Hf_2SeC$ | Theoretical Mode symmetry | IR $Hf_2TeB$ | Raman $Hf_2TeB$ | Theoretical Mode symmetry | IR $Hf_2TeC$ | Raman $Hf_2TeC$ |
| E1u | - | - | E1u | - | - | E1u | - | - | A2u | - | - | E1u | - | - | E1u | - | - |
| E1u | - | - | E1u | - | - | E1u | - | - | E1u | - | - | E1u | - | - | E1u | - | - |
| A2u | - | - | A2u | - | - | A2u | - | - | E1u | - | - | A2u | - | - | A2u | - | - |
| E2u | - | - | E2u | - | - | E2u | - | - | E2u | - | - | E2u | - | - | E2u | - | - |
| E2u | - | - | E2u | - | - | E2u | - | - | E2u | - | - | E2u | - | - | E2u | - | - |
| B1u | - | - | B1u | - | - | E2g | - | 104.6 | E2g | - | 111.7 | E2g | - | 90.1 | E2g | - | 95.1 |
| E2g | - | 113.9 | E2g | - | 118.8 | E2g | - | 104.6 | E2g | - | 111.7 | E2g | - | 90.1 | E2g | - | 95.1 |
| E2g | - | 113.9 | E2g | - | 118.8 | B1u | - | - | B1u | - | - | B1u | - | - | B1u | - | - |
| E1g | - | 131.5 | E1g | - | 138.2 | E1g | - | 133.1 | E1g | - | 144.4 | B2g | - | - | B2g | - | - |
| E1g | - | 131.5 | E1g | - | 138.2 | E1g | - | 133.1 | E1g | - | 144.4 | E1g | - | 134.8 | E1u | 139.2 | - |
| B2g | - | - | B2g | - | - | B2g | - | - | B2g | - | - | E1g | - | 134.8 | E1u | 139.2 | - |
| A1g | - | 194.1 | A1g | - | 198.8 | E1u | 168.6 | - | E2g | - | 176.2 | E1u | 137.0 | - | E1g | - | 142.8 |
| E2g | - | 233.0 | E2g | - | 250.3 | E1u | 168.6 | - | E2g | - | 176.2 | E1u | 137.0 | - | E1g | - | 142.8 |
| E2g | - | 233.0 | E2g | - | 250.3 | E2g | - | 169.5 | E1u | 176.4 | - | E2g | - | 151.7 | E2g | - | 159.7 |
| E1u | 239.4 | - | E1u | 251.0 | - | E2g | - | 169.5 | E1u | 176.4 | - | E2g | - | 151.7 | E2g | - | 159.7 |
| E1u | 239.4 | - | E1u | 251.0 | - | A1g | - | 190.5 | A1g | - | 205.5 | B2g | - | - | A2u | 179.9 | - |
| A2u | 295.5 | - | A2u | 309.2 | - | B2g | - | - | A2u | 218.0 | - | A1g | - | 181.3 | A1g | - | 194.2 |
| B2g | - | - | B2g | - | - | A2u | 215.4 | - | B2g | - | - | A2u | 182.8 | | B2g | - | - |
| A2u | 573.5 | - | B1u | - | - | A1u | 551.1 | - | B1u | - | - | B1u | - | - | A1u | - | - |
| E1u | 579.8 | - | A2u | 544.6 | - | A1u | 551.1 | - | A2u | 539.2 | - | E1u | 543.0 | - | A2u | 511.8 | - |
| E1u | 579.8 | - | E1u | 616.4 | - | A2u | 551.9 | - | E1u | 601.6 | - | E1u | 543.0 | - | E1u | 561.8 | - |
| E2u | - | - | E1u | 616.4 | - | E2u | - | - | E1u | 601.6 | - | A2u | 545.7 | - | E1u | 561.8 | - |
| E2u | - | - | E2u | - | - | E2u | - | - | E2u | - | - | E2u | - | - | E2u | - | - |
| B1u | - | - | E2u | - | - | B1u | - | - | E2u | - | - | E2u | - | - | E2u | - | - |

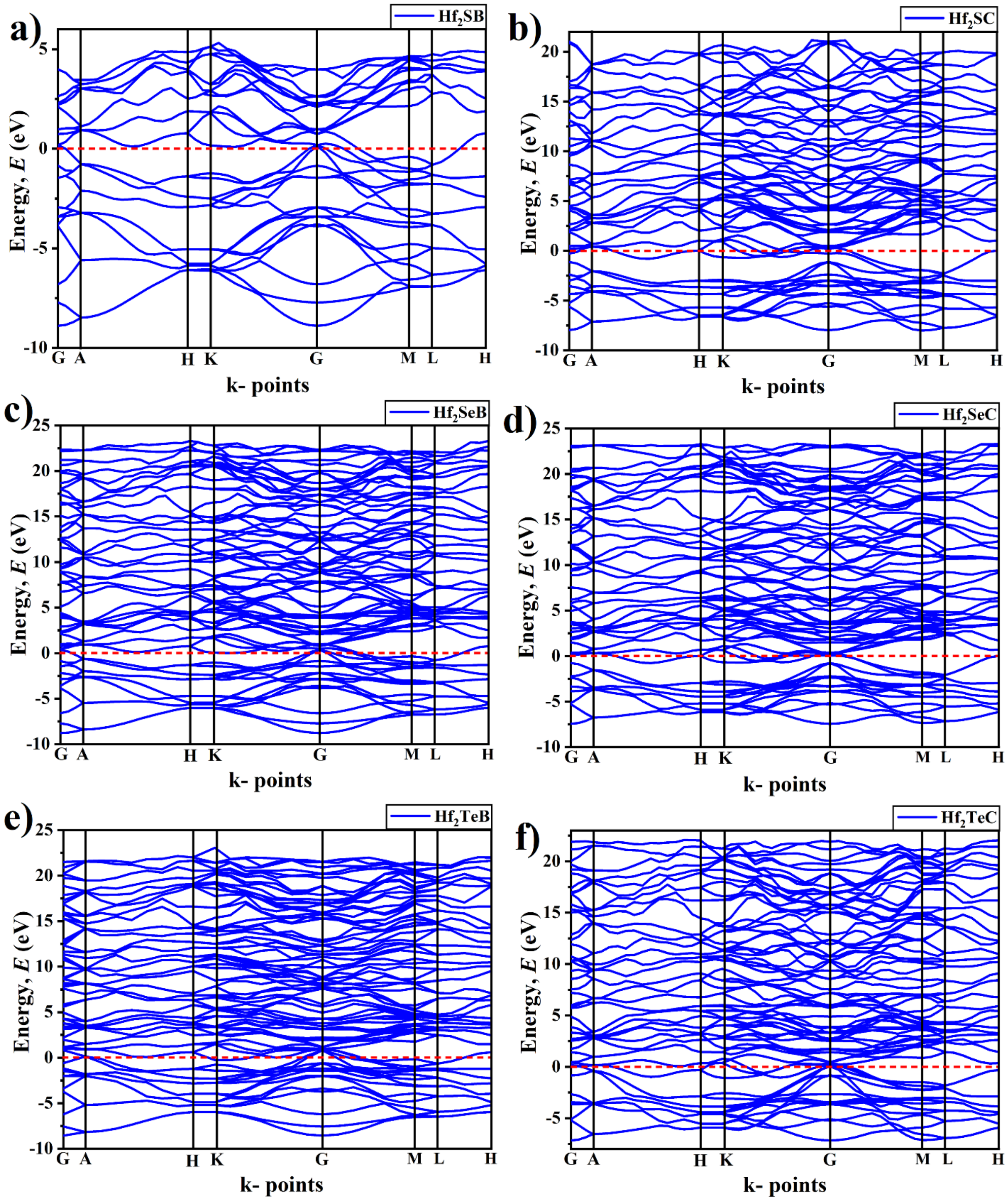

**Figure S1.** Calculated electronic band structure of a) $Hf_2SB$, b) $Hf_2SC$, b) $Hf_2SeB$, d) $Hf_2SeC$, e) $Hf_2TeB$, and f) $Hf_2TeC$ along the high-symmetry directions in the Brillouin zone. Red dotted line indicates the Fermi level ($E_F$).

**Corresponding Authors:** ashrafphy31@cuet.ac.bd, salehnaqib@yahoo.com